# The Infrared Imaging Spectrograph (IRIS) for TMT: Instrument Overview

James E. Larkin[*a], Anna M. Moore[b], Shelley A. Wright[c], James E. Wincentsen[b], David Anderson[e], Eric M. Chisholm[d], Richard G. Dekany[b], Jennifer S. Dunn[e], Brent L. Ellerbroek[d], Yutaka Hayano[f], Andrew C. Phillips[g], Luc Simard[e], Roger Smith[b], Ryuji Suzuki[f], Robert W. Weber[b], Jason L. Weiss[a], and Kai Zhang[h]

[a]Department of Physics and Astronomy, University of California, Los Angeles, CA 90095-1547;
[b]Caltech Optical Observatories,1200 E California Blvd., Mail Code 11-17, Pasadena, CA 91125;
[c]Center for Astrophysics & Space Sciences, University of California, San Diego, 9500 Gilman Drive, La Jolla, CA 92093
[d]Thirty Meter Telescope Observatory Corporation,100 West Walnut Street, Suite 300, Pasadena, CA, 91124
[e]Herzberg Institute of Astrophysics (HIA), National Research Council Canada, 5071 W Saanich Rd, Victoria, V9E 2E7;
[f]National Astronomical Observatory of Japan, 2-21-1 Osawa, Mitaka, Tokyo, 181-8588, Japan
[g]University of California Observatories, CfAO, University of California, 1156 High St., Santa Cruz, CA, 95064;
[h]National Astronomical Observatories, Nanjing Institute of Astronomical Optics and Technology, Chinese Academy of Sciences, No. 188 Bancang St., Nanjing, Jiangsu, 210042, China

## ABSTRACT

IRIS is a near-infrared (0.84 to 2.4 micron) integral field spectrograph and wide-field imager being developed for first light with the Thirty Meter Telescope (TMT). It mounts to the advanced adaptive optics (AO) system NFIRAOS and has integrated on-instrument wavefront sensors (OIWFS) to achieve diffraction-limited spatial resolution at wavelengths longer than 1 µm. With moderate spectral resolution ($R \sim 4000 - 8,000$) and large bandpass over a continuous field of view, IRIS will open new opportunities in virtually every area of astrophysical science. It will be able to resolve surface features tens of kilometers across Titan, while also mapping the most distant galaxies at the scale of an individual star forming region. This paper summarizes the entire design and capabilities, and includes the results from the nearly completed preliminary design phase.

## 1. INTRODUCTION

The Infrared Imaging Spectrograph (IRIS) is a fully cryogenic instrument being developed for first-light operation of the Thirty Meter Telescope (TMT)[1] and its Narrow Field Infrared Adaptive Optics System (NFIRAOS)[2]. IRIS combines a "wide field" imager[3,4] and an integral field spectrograph[5,6] both covering wavelengths from 0.84 µm to 2.4 µm. Three on-instrument wavefront sensors (OIWFS)[7] have been developed and prototyped at National Research Council Herzberg (NRC-H) and serve to track tip-tilt and focus errors that the laser guide star system is blind to, and to properly compensate low-order modes of field distortion using the multi-conjugate AO system . Both science channels are built around the latest 4K by 4K HgCdTe detectors[8] (H4RG-10 for the imager, H4RG-15 for the spectrograph) from Teledyne. The imager has 4 milliarcsecond (mas) pixels and a 2x2 assembly of detector arrays generating a total field of view of 34 x 34 arcsec (including small gaps). The spectrograph offers four plate scales ranging from 4 mas to 50 mas and can take up to 14,000 spectra simultaneously in a filled rectangular pattern. Figure 1 shows the delivered two arcminute field of view from the adaptive optics system and how it's utilized by the various subsystems.

---

[*] larkin@astro.ucla.edu; phone 1 310 825-9790; fax 1 310 206-2096

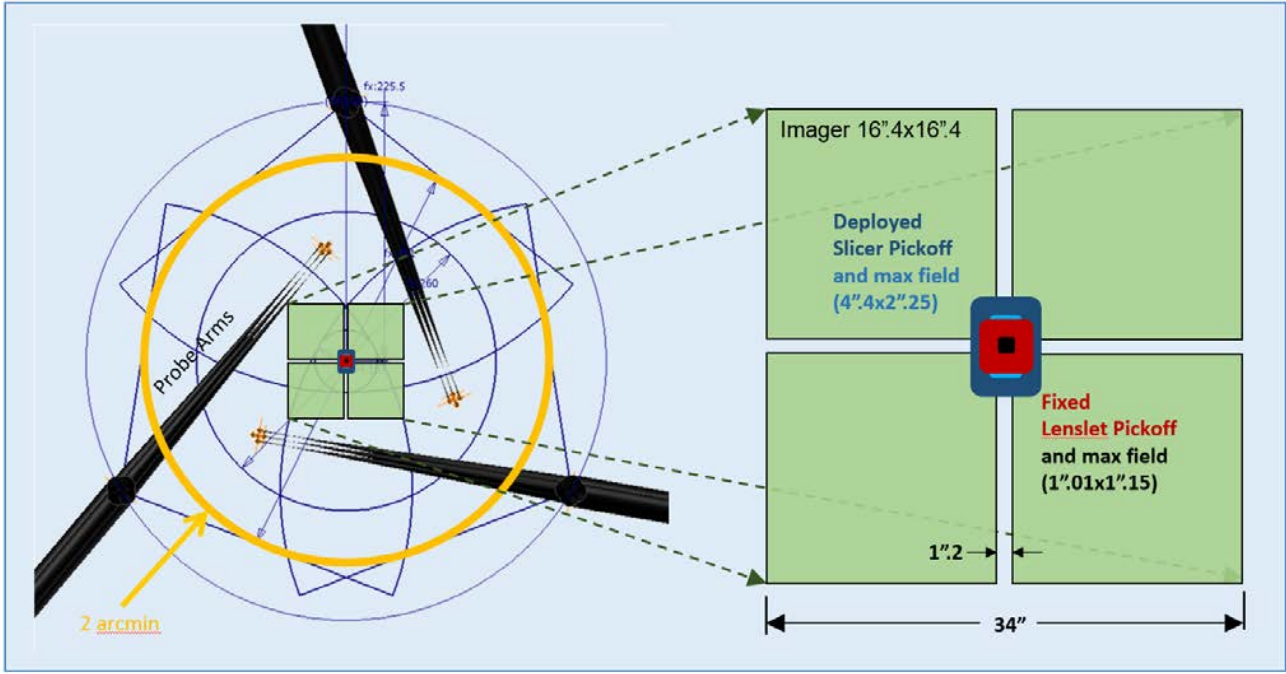

Figure 1. The division of the IRIS entrance field relative to the On-Instrument Wavefront Sensor probe arms. Entrance field is 2 arcmin in diameter and is patrolled by the 3 OIWFS probe arms. The green squares are the tiled 34" imager field of view. The concentric lenslet and image slicer IFU fields (four channels in total) are located at the center of the field as well to optimize the use of the best region of performance of NFIRAOS. The blue slicer fields are selected by a moving pickoff mirror and for lenslet operation, less of the imager field is vignette.

IRIS has been presented multiple times at previous SPIE meetings including overview papers in 2010[4] and 2014[5]. So this paper will concentrate on recent design changes especially regarding optical and mechanical changes. These include expanding the imager field of view to 34 arcseconds on each side using a 2x2 mosaic of Hawaii-4RG detectors and the use of the imager as a reimaging system for the spectrograph optics so they can share many components including atmospheric dispersion compensators. But it is still useful to give some of its overall capabilities here.

The science fields of view and spectral capabilities are given in Table 1 below. The slicer spectrograph divides the field into 88 slices that are reformatted into two long slits at the entrance to a collimator. For special modes where very long (between 2000 and 4000 elements) spectra are desired, one of the two long slits can be blocked reducing the field by half. Each slice has 45 spatial elements along its length producing fields of 88x45 or 44x45 (with mask) spatial elements in rectangular formats. The lenslet array spectrograph has a normal mode of 112x128 spatial elements and a secondary mode of 16x128 lenslets to increase the number of available spectral elements to between 2000 and 4000. The imager mode can be used in parallel with any of the spectrograph modes but through a common filter. The four imager detectors can also be independently read out and regions within them can serve as additional tip/tilt sensors.

**Table 1:** A top-level summary of IRIS capability

| Capability mode | Spatial sampling (mas) | Field of View (arcsec) | Resolution ($\lambda/d\lambda$) | Min/Max wavelength ($\mu m$) | Bandpass |
|---|---|---|---|---|---|
| Imager | 4 mas | 34 x 34 | Set by filter | 0.84-2.4 | 37 filters Variety of bandpasses |
| Slicer IFS | 50 mas | 4.4 x 2.25 | 4,000, 8,000 | 0.84-2.4 | 20%,10% |
| 88x45 Spaxels | 25 mas | 2.2 x 1.125 | | 0.84-2.4 | 20%,10% |
| *Slicer IFS* | 50 mas | 4.2 x 2.25 | 4,000-10,000 | 0.84-2.4 | 20%,10%, H+K |
| *44x45 Spaxels* | 25 mas | 1.1 x 1.125 | | 0.84-2.4 | 20%,10%, H+K |
| *Lenslet IFS* | 9 mas | 1.01 x 1.15 | 4,000 | 0.84-2.4 | 5% |
| *112x128 Spaxels* | 4 mas | 0.45 x 0.51 | | 0.84-2.4 | 5% |
| *Lenslet IFS* | 9 mas | 0.144 x 1.15 | 4,000-10,000 | 0.84-2.4 | 20%, H+K |
| *16x128 Spaxels* | 4 mas | 0.064 x 0.51 | | 0.84-2.4 | 20%, H+K |

## 1.1 TMT and the Narrow Field Infrared Adaptive Optics System

A fundamental motivation for the construction of extremely large telescopes like TMT is the ability to achieve unprecedented angular resolutions with a single aperture. Given the turbulent nature of the atmosphere, the diffraction limit can currently only be reached with adaptive optics. So the TMT team has always seen adaptive optics as a core capability that should be available at first light. The Narrow Field Infrared Adaptive Optics System (NFIRAOS) is an advanced multiconjugate adaptive optics system being designed in parallel with the telescope. NFIRAOS has two deformable mirrors and can use both natural and laser guide stars to produce a well corrected two arcminute field of view. The large beam and proximity to the telescope prohibits an internal field rotation system so instruments mount to one of three ports where they rotate about their beam axis. Since laser guide stars cannot be used to measure atmospheric tip/tilt or focus and because the use of two deformable mirrors can distort the field, each instrument must have its own sensors to monitor and maintain image distortion and field location. In IRIS we have three tip/tilt wavefront sensors that patrol the full field and simultaneously track reference stars in the near infrared. Figure 2 shows the IRIS instrument with its cable wrap using the "up-looking" port of NFIRAOS on the Nasmyth deck. This arrangement also provides IRIS a fixed gravity orientation.

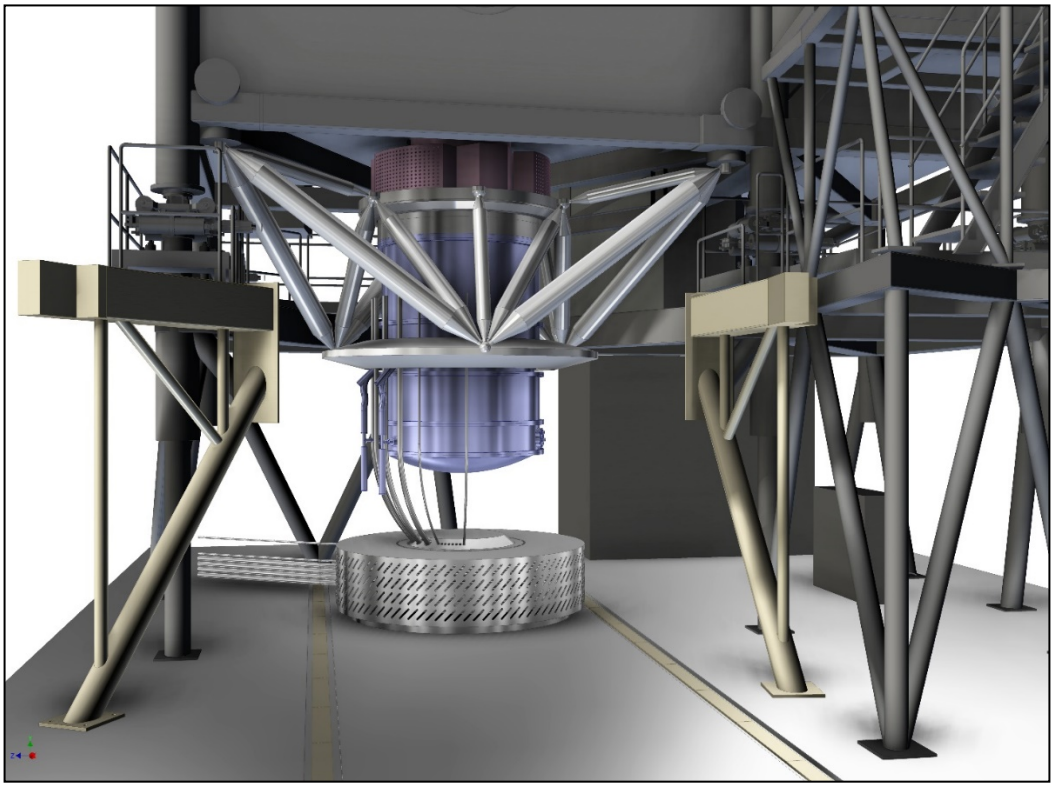

Figure 2. The entire IRIS instrument shown as it will appear mounted to the up-looking port of the NFIRAOS adaptive optics system. The silver cylinder below the blue dewar is the cable wrap to prevent damage to cables and services as the dewar rotates during operation. For scale, the blue dewar is 1.9 meters in diameter.

## 2. SCIENTIFIC JUSTIFICATION

We have a large international science team (24 members) from six countries continually working to refine IRIS's capabilities and motivate its construction. Our science team has investigated many science cases that TMT and IRIS will offer to a range of astronomical fields from the solar system to first light galaxies. Due to space limitations, we only present science cases which emphasize the power of simultaneous observations with the spectrograph and the wider format imager (quadruple the area on the sky) that has become our new baseline. These are *just a small sample* of IRIS and TMT science cases that highlight the uniqueness of combining these diffraction-limited data sets. Other cases can be found in a variety of publications including the recent overview by our Project Scientist, Shelley Wright[9].

### 2.1 Characterizing the first galaxies in the universe

The epoch of "first light" – between 300 and 900 million years after the Big Bang – is a key period in cosmic evolution for understanding the first seeds of stellar, supermassive black hole (SMBH), and galaxy formation. TMT and IRIS will explore this "First Light" epoch by revealing extremely faint star-forming galaxies and by studying the detailed physical characteristics of individual early galaxies. The *James Webb Space Telescope* (JWST) is designed to have the sensitivity and spatial resolution needed to find certain types of early (z > 7) galaxies. With a focus on broad-band and narrow-band imaging capabilities only at longer wavelengths ($\lambda > 1.6$ μm), JWST will find continuum objects at extreme redshifts, z > 8. The capabilities of TMT/IRIS will be extremely complementary, allowing the discovery of line-emitting (Lyman $\alpha$) populations that are, most likely, distinct from the populations that JWST will find. The most efficient mode of discovering these galaxies is likely a combination of blind and targeted narrow-band imaging with the IRIS imager. Because Lyman $\alpha$ emission does not penetrate a neutral medium, the clustering of Lyman $\alpha$ sources on Mpc scales reveals the growth of these bubbles throughout cosmic time. Thus, with narrow-band maps of large areas of sky, we will be able to constrain the luminosity function and the clustering of Lyman $\alpha$ sources and reveal extensive information about the nature of reionization.

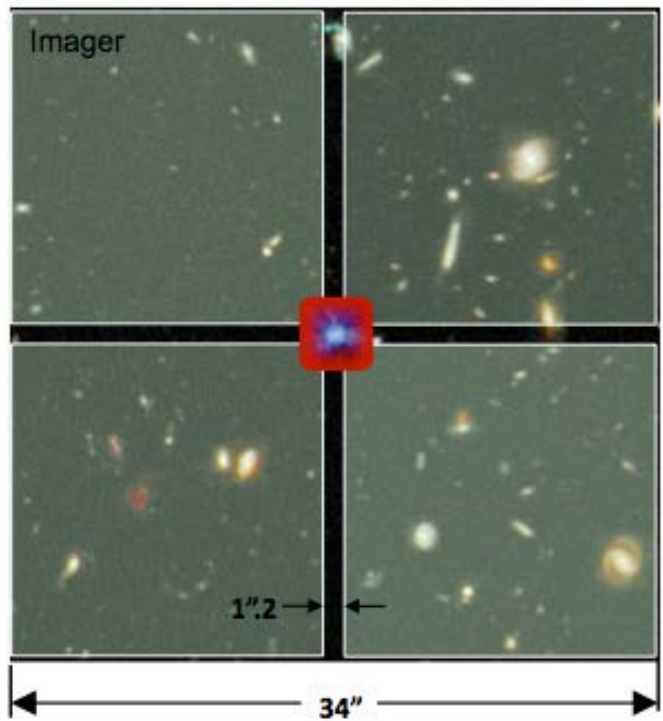

Figure 3. An illustration of the integral field spectrograph (IFS) and imager setup to conduct observations in the Hubble Deep Field that is to scale. The focal plane is tiled with four H4RG-10 arrays separated by a gap of 1.2". The IFS is on-axis (red) taking resolved spectra of a high-redshift galaxy, while simultaneous observations are being conducted using parallel imaging

field of view. A 20 minute integration time with the IRIS imager in its broad-band will be able to archive comparable depth to today's WFC3 exposures.

By targeting regions around large galaxies discovered by JWST or the gravitationally lensed regions around nearby massive galaxy clusters, IRIS can explore different portions of the Lyman $\alpha$ luminosity function. The majority of the first light (z >8) galaxies discovered in the coming years will be in the traditional deep extragalactic fields (i.e., Hubble Deep Field, Extended Groth Strip). As illustrated in Figure 3, the parallel mode of IRIS will be incredibly powerful allowing us to take deep resolved spectroscopic observations on a particular high-redshift galaxy while the imager makes simultaneous images of the wider field. TMT with the IRIS imager will be able to achieve observations of comparable depth as the Hubble Deep Fields in 20 minutes with significantly better spatial resolution.

## 2.2 Dynamics and stellar populations in the Galactic Center

One of the leading science cases for IRIS is to study the Galactic Center with unprecedented sensitivity and spatial resolution. The Galactic Center is the closest laboratory for studying the environments and fundamental physics of SMBHs and offers very unique science cases for TMT. Both the IRIS imager and IFS have been carefully designed to characterize the surroundings of the SMBH, SgrA*, at the Milky Way's center. If we can achieve our goal of relative astrometric accuracy of 30 μas we will be able to test General Relativity and probe the distribution of dark matter through orbital monitoring of the stars surrounding SgrA*. This level of astrometric accuracy can only be achieved with the precision reference frame provided by several maser sources with radio interferometric positions. Currently, the limited field of view of Keck and VLT observations means that the observations have to be dithered to capture these masers. The accuracy of these determined positions are limited by uncertainties and variability in camera distortion and severely limit the astrometric accruacy achieved at the Galactic Center today. Using the IRIS expanded imager capability (34"x34") we will be able to observe 8 maser sources in a *single* observation, which will greatly improve the astrometric accuracy in this field. This is imperative for measuring the orbits of current and new inner-arcsecond sources surrounding Sgr A* and exploring the fundamental physics of the SMBH. The expanded IRIS imager will also allow greater accessibility and ease of studying the stellar population at the Galactic Center (as illustrated in Figure 4). The young stellar population near SgrA* has puzzled astronomers, as young massive stars should have difficulty forming in close proximity to a SMBH.

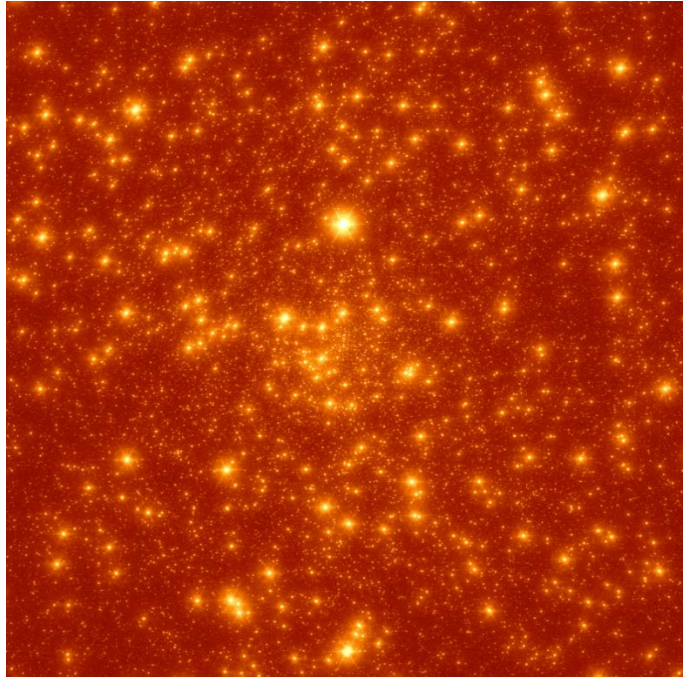

Figure 4. Simulation of an IRIS imager observation using K (2.2 μm) broadband of the Galactic Center with the expandable field of view of 32.8" x 32.8". In a single shot using the 0.004" plate scale the IRIS imager will be able to capture the inner-arcsecond sources, Sgr A*, the stellar population, and essential astrometric reference frame sources of 8 masers. This simulated observation represents only 20 seconds of exposure time. Based on the extrapolated stellar population IRIS would be able to detect 500,000 stars down to K < 25 mag (T. Do, IRIS science team).

# 3. MAJOR SUBSYSTEMS

IRIS has three distinct optical units that occur sequentially after the adaptive optics focal plane. At the top of the instrument are the on-instrument wavefront sensors (OIWFS) which monitor up to three stars across a 2 arcmin region of the AO corrected field. Their primary purposes are to remove image distortion and provide tip/tilt and focus tracking for IRIS. The first science portion is an on-axis "wide field" imager with a 34 x 34 arcsec field of view with 4 mas pixels. The integral field spectrographs use two small pick-off mirrors just before the imager detectors to observe small central fields with selectable plate scales between 4 mas (Nyquist sampling at 1 micron) and a coarsest platescale of 50 mas (maximized sensitivity on low surface brightness sources). To achieve this large range of scales, two different techniques are used and we refer to this as a "hybrid" design. The two finest scales (4 and 9 mas) use a lenslet-based optical design to maximize field of view (112x128 spatial samples) and achieve the highest possible image quality. The coarsest two scales (25 and 50 mas) use a slicer-based design to provide more flexible spectral formats and greater spectral bandwidth (spectra up to 2000 pixels in length even with full field). The four IFU plate scales are concentric with each other and simple two position stages are used to divert the light into the separate optical paths with high positional accuracy. A coincidence of the slicer and lenslet design parameters makes an F/4 camera optimal for formatting the light from the gratings onto the science detector. So after the pick-off mirrors in the imager, the slicer and lenslet beams are separated and then light is recombined at the grating turret with a common set of gratings, camera optics and detector. This sharing of most of the optical elements leads to major cost savings in the instrument and actually would not be possible if a single slicer or lenslet had been attempted for all four plate scales.

## 3.1 On Instrument Wavefront Sensor

The NFIRAOS adaptive optics system will often use laser guide stars and perform a multi-conjugate analysis of the atmosphere. This allows it to correct stellar images over a full 2 arcminute field, but the LGS AO system itself is blind to several of its own corrective modes. In particular, field distortion, tip/tilt and focus must be determined independently using real stars and natural guide star sensors. For full correction of these effects, three stars must be monitored in the corrected field of view. Given the density of stars on the sky, achieving full correction over at least half of the sky (or equivalently for half of anyone's targets of interest) requires using very faint stars where AO correction is needed to improve their contrast against the background light. As a result, IRIS is designed with three pick-off arms which patrol the AO corrected field of view of NFIRAOS and which will use infrared wavefront sensors to gain from the high Strehl ratios in the infrared. To further improve the sensitivity, the sensors will integrate over a very wide wavelength range (1.15-2.3 microns). The sensors are cryogenic infrared detectors cooled to liquid nitrogen temperatures. To sufficiently reduce the thermal background the optical components including probe arms must be cooled to roughly -30 C.

Since the OIWFS serves as the direct measurement of stars on the sky, its performance is often the dominant factor in the astrometric performance of the system. In some cases we also envision using guide regions of the imager detectors as tip/tilt sensors to further improve astrometric performance. But in general, if one of the probes is misplaced or makes a poor centroid measurement, then the wavefront sensors will command the AO system to move that star to the wrong position, which will both distort and blur the image delivered to the science cameras. Since there is differential atmospheric dispersion between the LGS wavefront sensors wavelength range, the OIWFS wavelength range, and the science wavelength, the stellar images are also moving differentially in real time from the optical position of the stars, and with respect to the science target. As a result, the probe arms must move very accurately (typical rms of better than 4 microns) and must track well as stars transition across pixel boundaries. The probe arms have not changed significantly in optical design in the past two years and we refer interested readers to past publications[7].

## 3.2 "Sequential Design"

One of the most recent and impactful optical decisions has been to use what we term a "sequential design" in which the science light first passes through all of the imager optics and then small pick-off mirrors are used to divert the light into reimaging optics that feed the spectrographs. As we discuss in some detail below, this has a variety of advantages including putting all science fields on the region of best AO correction, minimizing non-common path wavefront errors between science legs and greatly reducing variable image distortion and throughput variations from very small ADC prisms in earlier separate spectrograph designs. This choice also significantly reduces the number of cryogenic mechanisms within the instrument and potentially allows the imager detector to serve as an additional tip/tilt reference sensor for some spectrographic modes. As we'll discuss in the mechanical section, the sequential design also simplifies the integration and testing of the separate subsystems prior to full integration.

### 3.3 Hybrid Spectrograph

As we have described the spectrograph uses both a lenslet and a slicer integral field unit to cover the four plate scales. There are many motivations for this and they have been described in several publications[5, 6]. One of the biggest benefits is the sharing of many components including all filters, the ADC, Lyot stop, gratings, camera optics and Hawaii-4RG15 detector. In particular the camera optics are fast with a large field angle and perhaps the most difficult optical system in the instrument. The large field angle also led us to perform a set of comparison measurements between ruled and holographic gratings. In general, ruled gratings have more variation as a function of wavelength compared to a holographic grating, but less variation with output or beta angle. Since we already plan to use $1^{st}$ order gratings to avoid order overlap over the large field angle and because our longest primary bandpasses are 20%, our suspicion was that ruled gratings would have less variation for our design. This was tested and published by our team and we indeed found that over our ranges of field angle and bandpasses ruled diffraction gratings had a higher overall efficiency compared to volume phase holographic gratings. Two companies were used for each type of grating and were tested at infrared wavelengths with our selected field angles.[10, 11] So the beams from the lenslet and slicer collimators are directed to strike a common rotating turret of ruled diffraction gratings at the same incident angles and thus are directed into the common camera optics to the detector.

### 3.4 Lenslet Array Integral Field Spectrograph

With any adaptive optics science instrument one of the greatest challenges is the preservation of the excellent image quality and corresponding low wavefront error. Spectrographs in particular typically have more optical elements and assemblies and contain components that are often much higher in surface irregularity such as diffraction gratings. Preserving wavefront quality is particularly challenging at the finest platescales and shortest wavelengths. In an integral field spectrograph at fine plate scales, it is also often a challenge to obtain a large enough field of view to ensure the full PSF is captured and to cover scientifically interesting objects in individual exposures. For both of these problems integral field spectrographs based on lenslet arrays offer significant advantages. The physical arrays are typically created with lithographic processes that are easy to expand to large numbers of elements. These arrays are also utilized before the spectrograph collimator, grating and camera to sample the image plane. So only aberrations prior to the lenslet array contribute directly to wavefront error and several spectrographs like OSIRIS at Keck have demonstrated non-common path wavefront errors below 30 nm rms.[12] In the case of IRIS, the imager optics are a contributing factor to the spectrograph wavefront error, but these are precision optics where field curvature remains among the largest terms. Since the spectrograph uses only a tiny portion of the field, the curvature doesn't affect image quality. More importantly the wavefront error can be partially corrected by NFIRAOS. The ability to simultaneously correct much of the residual wavefront error in the imager and the spectrograph is a major advantage of the sequential design.

So for the reasons just stated, along with strong optical compatibility with the slicer optics used for coarser scales, a lenslet array was chosen as the integral field unit for the two finest plate scales of 4 and 9 mas. The selected lenslet array has a square pattern and a pitch of 350 microns per lenslet. The array is placed in the focal plane of simple reimaging optics and the pupil produced by each lenslet is typically 32 times smaller than its own diameter including the effects of pupil diffraction. This compression is sufficient to allow 16 spectra to interleave together in the projected height of an individual lenslet. So a basic pattern of 16x16 lenslets is used as a building block for the full field. A 7x8 pattern of these basic blocks are then used to create the 112x128 spatial samples. Each lenslet's spectrum can extend for the projected length of a 16x16 block or approximately 512 pixels. In the dominant modes where the spectral resolution is 4000, this corresponds to a 5% bandpass in each individual exposure. A mask can be used to restrict the field to 16x128 lenslets giving each lenslet an entire row of the detector for its spectra. This field restricted mode allow spectra almost up to 4000 pixels in length to be produced and could include broadband spectra up to resolutions of 10,000, or much broader spectra (e.g. H+K) at resolutions of 4000.

### 3.5 Slicer Integral Field Spectrograph

At coarse platescales, wavefront error is not a significant concern, at least not errors that blur the core. Most objects of interest also become much easier to cover and fields of 100x100 elements are not needed. It becomes more optimal to have slightly fewer spatial samples but longer spectra to improve efficiency. Mirror slicers have the advantage over lenslets that they can reformat the focal plane into roughly linear strips (referred to as long slits) that do not cause overlap issues on the detector. So they over more flexible spectral formats and are less subject to spectral crosstalk issues of lenslets. Wavefront quality is sacrificed, but not in a way that affects coarse modes. So for the coarse scales of 25 and 50 mas, IRIS will use 88 flat mirrors arranged in a stack in the focal plane as the integral field unit. Pairs of spherical mirrors are used to reorganize the 88 resulting beams into the linear feeds to the spectrograph portion. The spherical mirrors also serve to de-magnify the slicer facets so the resulting linear inputs are six times shorter in physical length than the original facets aligned end-to-end. The two long slits are directed onto the left and right sides of the detector, respectively. This allows every spatial element to utilize approximately 2000 spectral pixels. A benefit of the slicer design is that adjacent spectra have nearly identical wavelengths next to each other making data reduction easier and cross contamination less of a problem. Since the last review we have added a masking mechanism that allows us to block one of the linear-feeds to the spectrograph cutting the field in half (44 slices). In this mode, like the narrow lenslet mode, each spatial location can produce a spectrum 4000 pixels in length and could include broadband spectra up to resolutions of 10,000, or much broader spectra (e.g. H+K) at resolutions of 4000.

### 3.6 Imager

Wavefront error and throughput are the driving concerns of the diffraction limited imaging camera in IRIS. During the conceptual design phase, wavefront error worries led to a design utilizing all refractive elements like apochromatic triplets and achromatic doublets. In part the focus on wavefront error was due to the inability to use the AO system to correct non-common path errors since those would be seen in the spectrograph. So the imager needed a raw wavefront error below 30 nm rms and refractive elements seemed the best strategy to achieve this. With the new goal of a much larger field (34" compared to 17") and the sequential design's ability to image sharpen both imager and spectrograph a thorough comparison of design options has led us to use all reflective three mirror anastigmats (TMA) for both the collimator and camera optics. They can support the larger field angles and don't reinforce the field curvature coming from the adaptive optics system like off-axis parabolas would. The all reflective design also gives better throughput than a refractive design especially at the longest wavelength of the system.

The imager has an extensive region of collimated space where a Lyot stop, atmospheric dispersion prisms and a stack of five filter wheels are located. As shown in the mechanical section below, 75 filter positions are provided both for spectroscopy and imaging needs. The Lyot stop is serrated and matched to the telescope aperture and actively tracks the primary rotation. The ADC prisms[13] are removable to maximize throughput at high elevations when differential dispersion is tolerable. For more details on the imager design see the recent papers in the SPIE.[3,4]

## 4. SCIENCE DEWAR MECHANICAL DESIGN

IRIS is a diffraction limited instrument, so one might naively assume that the instrument would be of comparable size to similar instruments on smaller telescopes. And indeed existing AO cameras would function well at TMT. But, the diffraction limit is 3-4 times smaller than with existing 8-10 meter telescopes and the fields of view of those cameras would be 3-4 times smaller. Similarly the field of correction of the multiconjugate AO system encourages us to maintain or even increase the field of view of the science instrument at the finer platescale. With new larger format infrared detectors being available, this can be achieved, but only with larger and more complex optical systems and corresponding large cryogenic vacuum chambers. The vacuum chamber for IRIS is shown in Figure 5. It is 1.9 meters in diameter and 2.9 meters tall. It is supported near the top from a rotational bearing and in operation rotates about its cylindrical axis.

Large cryostats produce their own set of dilemmas such as long thermal cycle times. So the mechanisms and components that go inside must be extremely reliable to prevent minor failures which would lead to month-long losses of observing time. Wherever possible we are building the cryostat and internal components on the heritage of several previous cryogenic instruments including the Keck instruments OSIRIS[12] and MOSFIRE[14]. Major subsystems will be fully tested in smaller chambers prior to integration in the full IRIS vessel including 10 year simulated lifetime tests of mechanisms identical in nature to delivered components.

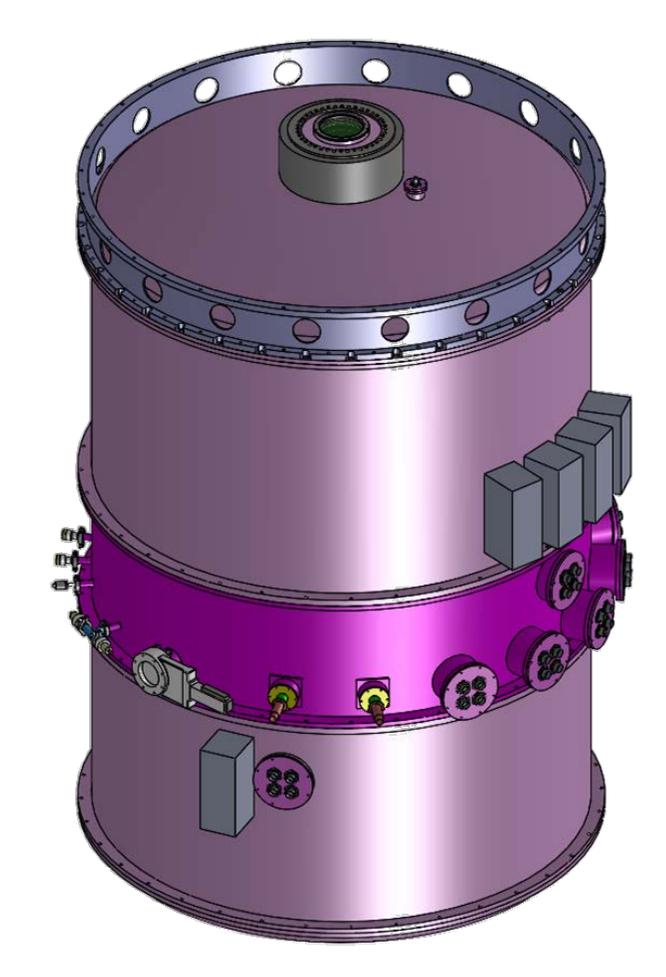

Figure 5 – Rendering of the current vacuum vessel for the cryostat. The diameter of the flanges is 1.9 meters and the overall height of what is shown here is 2.9 meters. To simplify integration and access to the imager in the upper section and the spectrograph in the lower section, a middle services ring (darker purple) contains most of the exterior connections including cryogenic helium gas flow, motor control, environmental monitoring, detector control, and data retrieval.

The sequential nature of the design led to a major simplification in that the imager and the spectrograph largely reside in separate halves of the dewar volume. We decided to actually divide the entire cryostat along the boundary and allow the upper and lower portions (shown in the left and right panels of Figure 6) to function as autonomous dewars for the integration of the imager and spectrograph, respectively. To combine the two systems together, the lower endcap of the imager and upper endcap of the IFS, along with a duplicate services ring are discarded, and the two systems are brought together.

The imager will be assembled and tested as an integrated assembly at the National Astronomical Observatory of Japan (NAOJ). Figure 7 shows the optomechanical assembly with cold plate, and shielding removed. As described above, the long collimated beam allows for large and complex mechanisms including the 75 position filter wheel shown in the right panel of Figure 7. Similarly the spectrograph will be assembled at Caltech and UCLA. Figure 8 shows portions of the spectrograph assembly that resides in the lower section of the overall cryostat. Again, a dominant feature is a large complex mechanism. In this case it's the grating turret with a diameter of 0.8 meters and containing 14 gratings.

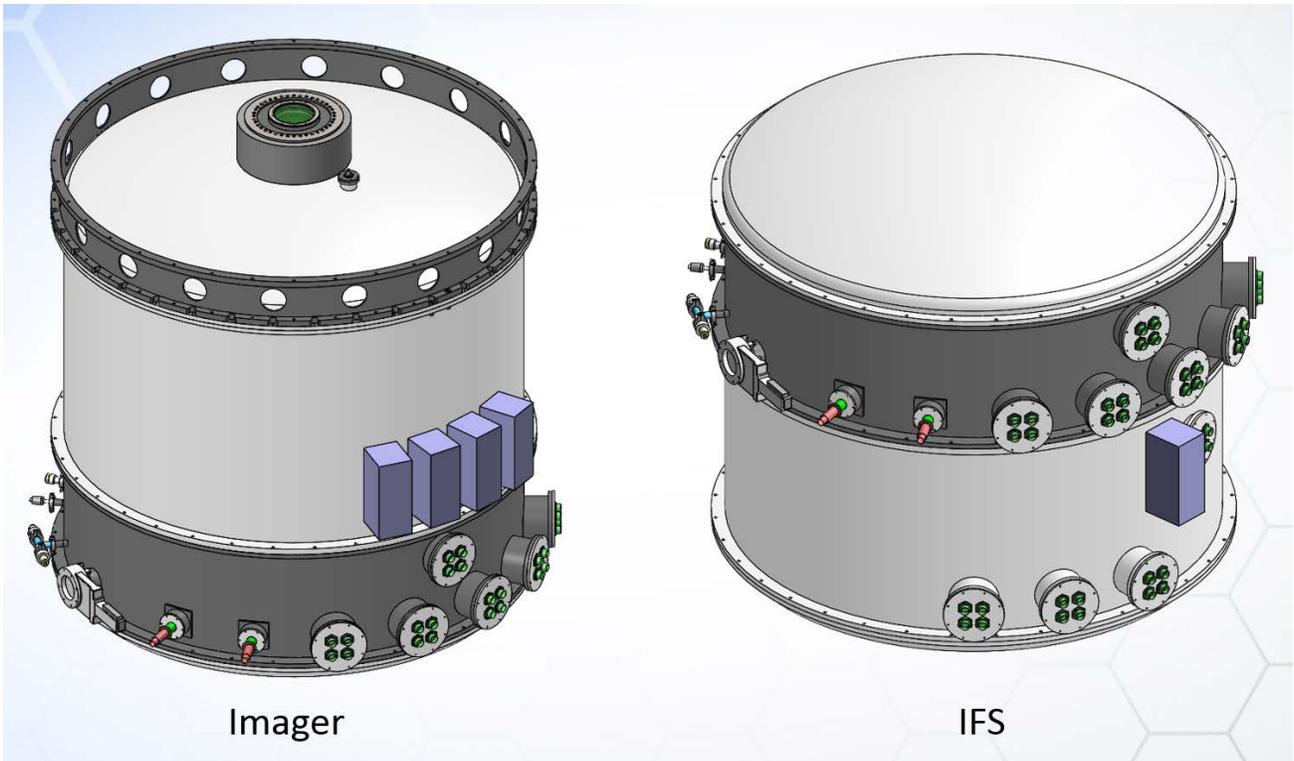

Figure 6 – Integration of a large international instrument poses its own set of challenges. The sequential nature of the design led to a major simplification in that the imager and the spectrograph largely reside in separate halves of the dewar volume. We decided to actually divide the entire cryostat along the boundary and allow the upper and lower portions (here left and right) to function as autonomous dewars for the integration of the imager and spectrograph (IFS) respectively.

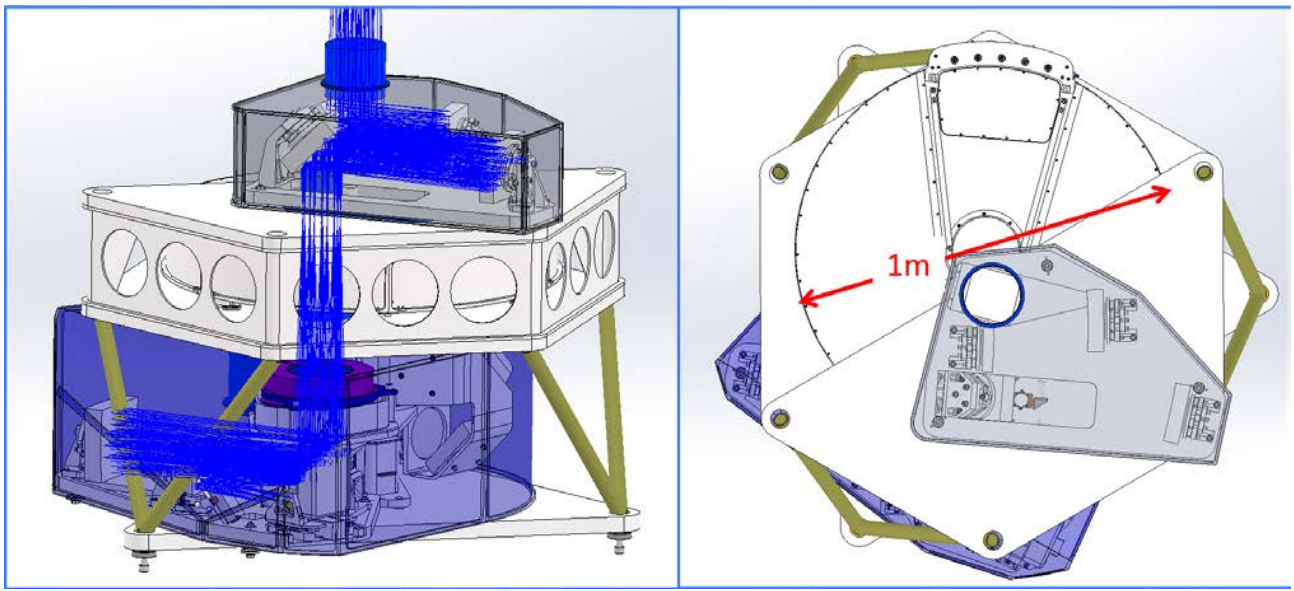

Figure 7 – The left panel shows the mechanical layout of the Imager optical system as viewed from the side. The right panel is a top down (as seen from NFIRAOS) view illustrating the large size of the filter wheel.

### 4.1 Cryogenic Mechanisms

As with any cryogenic instrument, especially at the size scale of IRIS, mechanisms should be minimized and must be extremely robust (typically with 10-year demonstrated life). For the science dewar, where mechanisms will likely operate between 77 and 120 Kelvin, the following mechanisms are currently being designed:

Imager Mechanisms:

- Atmospheric dispersion compensation system – two sets of counter-rotating prisms – continuous motion
- Rotating pupil mask – simple rotation but with continuous motion
- Filter mechanism with 5 wheels and a total of 75 potential filter locations. See Figure 7
- Pupil viewing selection mirror – two position stage
- Detector focus – micron level motion control

Spectrometer Mechanisms:

- Slicer pick-off mirror to divert light from the center of the field to the slicer stack.
- Plate scale selector. A single 2-position stage selects between 4 and 9 mas scales for the lenslet and between 25 and 50 mas in the slicer spectrographs.
- Lenslet spectrograph uses a slit scanning mechanism for calibration and for reducing the field for broad band modes.
- Slicer spectrograph uses a slide to mask off half of the field to allow for very extended spectral modes.
- Grating turret with 14 positions – requires micron level repetition
- Detector focus – micron level motion control

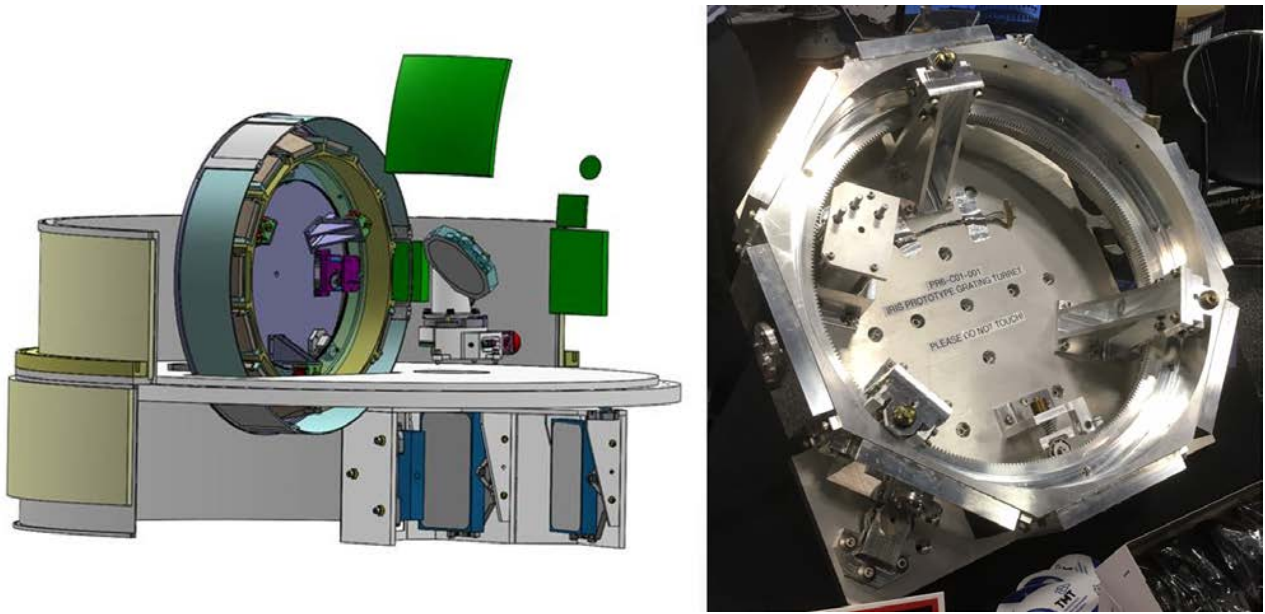

Figure 8 - The left panel shows a portion of the spectrograph section of the dewar showing the nested packaging of the three mirror anastigmats and the large grating turret. Thankfully the hybrid spectrograph design allows us to use one large rotating grating for all four plate scales and all spectral modes. The current grating turret has 14 positions. The right panel shows a prototype grating stage with 8 positions that fits inside the largest test chamber at UCLA for cryogenic testing.

## 5. SUMMARY

We have presented an update to the design of the IRIS instrument being developed for the Thirty Meter Telescope. We have optical and mechanical designs for all components and subsystems and have prototyped the most difficult elements such as the OIWFS probe arms and grating turret. The IRIS project is currently at the end of its preliminary design phase with a complete instrument review scheduled in November, 2016. Although site selection for the TMT has caused unexpected delays to the project, IRIS still needs to be completed rapidly. As a first light system that requires the adaptive optics system our integration phase requires us to be complete as a science instrument at least two years before telescope operation begins. So we currently anticipate starting a 2 year final design phase after the upcoming review followed by a 2 year fabrication phase, a 26 month integration phase and an 18 month integration and verification phase ending with instrument readiness in August, 2024. The following list gives our anticipated future milestones leading up to instrument readiness.

- December, 2016           Preliminary Design Phase Ends
- Jan 2017-Dec 2018        Final Design Phase
- Jan 2019-Dec 2020        Fabrication, Assembly and Subsystem Testing
- Jan 2021-Feb 2023        Integration – OIWFS goes to work with NFIRAOS
- Mar 2023-Aug 2024        Assembly Integration and Verification at the Telescope

## Acknowledgments

The TMT Project gratefully acknowledges the support of the TMT collaborating institutions. They are the California Institute of Technology, the University of California, the National Astronomical Observatory of Japan, the National Astronomical Observatories of China and their consortium partners, the Department of Science and Technology of India and their supported institutes, and the National Research Council of Canada. This work was supported as well by the Gordon and Betty Moore Foundation, the Canada Foundation for Innovation, the Ontario Ministry of Research and Innovation, the Natural Sciences and Engineering Research Council of Canada, the British Columbia Knowledge Development Fund, the Association of Canadian Universities for Research in Astronomy (ACURA), the Association of Universities for Research in Astronomy (AURA), the U.S. National Science Foundation, the National Institutes of Natural Sciences of Japan, and the Department of Atomic Energy of India.

## REFERENCES

[1] Sanders, G. H., "The Thirty Meter Telescope (TMT): An International Observatory," Journal of Astrophysics and Astronomy, 34, 2, 81-86 (2013)
[2] Glen Herriot ; David Andersen ; Jenny Atwood ; Corinne Boyer ; Peter Byrnes ; Kris Caputa ; Brent Ellerbroek ; Luc Gilles ; Alexis Hill ; Zoran Ljusic ; John Pazder ; Matthias Rosensteiner ; Malcolm Smith ; Paolo Spano ; Kei Szeto ; Jean-Pierre Véran ; Ivan Wevers ; Lianqi Wang ; Robert Wooff, "NFIRAOS: first facility AO system for the Thirty Meter Telescope", Proc. SPIE, 9148, 914810 (2014).
[3] Suzuki, R., Usuda, T., Crampton, D., Larkin, J., Moore, A., Phillips, A., and Simard, L., "The infrared imaging spectrograph (IRIS) for TMT: imager design," Proc. SPIE, 7735-42 (2010).
[4] Toshihiro Tsuzuki, Ryuji Suzuki, Hiroki Harakawa, Bungo Ikenoue, James Larkin, Anna Moore, Yoshiyuki Obuchi, Sakae Saito, Fumihiro Uraguchi, James Wincentsen, Shelley Wright, Yutaka Hayano, "The Infrared Imaging Spectrograph (IRIS) for TMT: Optical design of IRIS imager with 'Co-axis double TMA'," Proc. SPIE, 9908-386.
[5] James E Larkin, Anna M Moore, Elizabeth J Barton, Brian Bauman, Khanh Bui, John Canfield, David Crampton, Alex Delacroix, Murray Fletcher, David Hale, David Loop, Cyndie Niehaus, Andrew C Phillips, Vladimir Reshetov, Luc Simard, Roger Smith, Ryuji Suzuki, Tomonori Usuda, Shelley A Wright, "The infrared imaging spectrograph (IRIS) for TMT: instrument overview", Proc. SPIE, 7735 (2010).
[6] Anna M Moore, James E Larkin, Shelley A Wright, Brian Bauman, Jennifer Dunn, Brent Ellerbroek, Andrew C Phillips, Luc Simard, Ryuji Suzuki, Kai Zhang, Ted Aliado, George Brims, John Canfield, Shaojie Chen, Richard Dekany, Alex Delacroix, Tuan Do, Glen Herriot, Bungo Ikenoue, Chris Johnson, Elliot Meyer, Yoshiyuki Obuchi, John Pazder, Vladimir Reshetov, Reed Riddle, Sakae Saito, Roger Smith, Ji Man Sohn, Fumihiro Uraguchi, Tomonori

Usuda, Eric Wang, Lianqi Wang, Jason Weiss, Robert Wooff, "The infrared imaging spectrograph (IRIS) for TMT: spectrograph design", Proc. SPIE, 9147 (2014).
[7] Jennifer Dunn, Vladimir Reshetov, Jenny Atwood, John Pazder, Bob Wooff, David Loop, Leslie Saddlemyer, Anna M Moore, James E Larkin, "On-instrument wavefront sensor design for the TMT infrared imaging spectrograph (IRIS) update," Proc. SPIE, 9147 (2014).
[8] Majid Zandian et al. "Performance of science-grade HgCdTe H4RG-15 image sensors", Proc. SPIE, 9915-13 (2016)
[9] Wright, S. A., et al., "The Infrared Imaging Spectrograph (IRIS) for TMT: latest science cases and simulations", Proc. SPIE, 9905 (2016).
[10] Elliot Meyer, Shaojie Chen, Shelley A Wright, Anna M Moore, James E Larkin, Luc Simard, Jerome Marie, Etsuko Mieda, Jacob Gordon, "The infrared imaging spectrograph (IRIS) for TMT: reflective ruled diffraction grating performance testing and discussion", Proc. SPIE, 9147 (2014).
[11] Shaojie Chen, Elliot Meyer, Shelley A Wright, Anna M Moore, James E Larkin, Jerome Maire, Etsuko Mieda, Luc Simard, "The infrared imaging spectrograph (IRIS) for TMT: volume phase holographic grating performance testing and discussion", Proc. SPIE, 9147 (2014)
[12] Larkin, James; Barczys, Matthew; Krabbe, Alfred; Adkins, Sean; Aliado, Ted; Amico, Paola; Brims, George; Campbell, Randy; Canfield, John; Gasaway, Thomas; Honey, Allan; Iserlohe, Christof; Johnson, Chris; Kress, Evan; LaFreniere, David; Lyke, James; Magnone, Ken; Magnone, Nick; McElwain, Michael; Moon, Juleen; Quirrenbach, Andreas; Skulason, Gunnar; Song, Inseok; Spencer, Michael; Weiss, Jason; Wright, Shelley, "OSIRIS: a diffraction limited integral field spectrograph for Keck", Proc. SPIE, 6269 (2006).
[13] Phillips, Andrew, et al., "The Infrared Imaging Spectrograph (IRIS) for TMT: the ADC optical design", Proc. SPIE, 9908-373 (2016).
[14] McLean, I. S., Steidel, C. C., Epps, H. W., Matthews, K., and Adkins, S. M., "Design and development of MOSFIRE: the multi-object spectrometer for infrared exploration at the Keck Observatory", Proc. SPIE, 7735-49 (2010).